\documentclass[aps,prl,twocolumn,superscriptaddress,english]{revtex4-2}
\usepackage[utf8]{inputenc}

\usepackage{amsmath}
\usepackage{amssymb}
\usepackage{graphicx}
\usepackage{xcolor}
\usepackage{siunitx}
\usepackage{lipsum}

\usepackage{lineno}

\usepackage{hyperref}

\newcommand{\Tc}{$T_\text{c}$}
\newcommand{\ep}{$ep$}

\newcommand{\sh}{H$_3$S}

\newcommand{\cabh}{Ca(BH$_4$)$_2$}

\makeatletter
\def\@fnsymbol#1{\ensuremath{\ifcase#1\or \dagger\or *\or \ddagger\or
   \mathsection\or \mathparagraph\or \|\or **\or \dagger\dagger
   \or \ddagger\ddagger \else\@ctrerr\fi}}
\makeatother

\begin{document}

\author{Simone Di Cataldo}\email{simone.dicataldo@uniroma1.it}
\affiliation{Dipartimento di Fisica, Sapienza Universit\`a di Roma, 00185 Roma, Italy} 
\affiliation{Institute of Theoretical and Computational Physics, Graz University of Technology, NAWI Graz, 8010 Graz, Austria}
\author{Lilia Boeri} \email{lilia.boeri@uniroma1.it}
\affiliation{Dipartimento di Fisica, Sapienza Universit\`a di Roma, 00185 Roma, Italy} 
\affiliation{Centro Ricerche Enrico Fermi, Via Panisperna 89 A, 00184 Rome, Italy}

\title{Metal Borohydrides as high-\Tc{} ambient pressure superconductors}

\date{\today}
\begin{abstract}
The extreme pressures required to stabilize the recently discovered \textit{superhydrides} represent a major obstacle to their practical application. In this paper, we propose a novel route to attain high-temperature superconductivity in hydrides at ambient pressure, by doping
commercial metal borohydrides.
Using first-principles calculations based on Density Functional Theory and Migdal-\'{E}liashberg theory, we demonstrate that in \cabh{} a moderate hole doping of 0.03 holes per formula unit, obtained through a partial replacement of Ca with monovalent K, is sufficient to achieve \Tc{}'s as high as 110 K. The high-\Tc{} arises because of the strong electron-phonon coupling between the B-H $\sigma$ molecular orbitals 
 and bond-stretching phonons.
Using a random sampling of large supercells to estimate the local effects of doping, we show
that the required doping can be achieved without significant disruption of the electronic structure and at moderate energetic cost.
 Given the wide commercial availability of metal borohydrides, the ideas presented here can find prompt experimental confirmation. If successful, the synthesis of high-\Tc{} doped borohydrides will represent a formidable  advancement towards technological exploitation of conventional superconductors.
\end{abstract}

\maketitle


\section{Introduction}
Since the discovery of superconductivity with a critical temperature (\Tc{}) of 203 K at 150 GPa in \sh{} \cite{Eremets_Nature_2015_SH3, Eremets_NatPhys_2016_SH3}, hydrogen-rich superconductors have 
revolutionized the landscape of superconductivity research; After \sh{}, many other superhydrides with \Tc's{} close to, or even above, room temperature have been found \cite{Hemley_PRB_2018_LaH, Hemley_PRL_2019_LaH, Eremets_Nature_2019_LaH, Dias_Nature_2020_CSH, Ma_PRL_2022_CaH6, Oganov_MatToday_2020_ThH, Kong_NatComm_2021_YHx, Oganov_AdvMat_2021_YH6, Oganov_PRL_2021_CeH, Grockowiak_Frontiers_2022},
but the extreme synthesis pressures represent an insurmountable obstacle to any practical application.

On the other hand, an increasing demand exists for new materials 
enabling superconductor-based technologies: for most large-scale
applications, synthesizability at ambient pressure is 
a strict requirement, whereas the threshold for \Tc{}
is sensibly lower than ambient temperature, but is dictated
by the need to maintain a robust superconducting state under liquid nitrogen cooling. \cite{Pickard_AnnRevCMP_2020_review}

The spectacular success of computational methods for superhydrides
raises the hope that these techniques may accelerate
the identification of suitable materials.\cite{Boeri_JPCM_2019_viewpoint,Boeri_PhysRep_2020_review}
%
The first predictions which have started to
appear in literature follow essentially two different routes
to boost \Tc{} within the conventional (\ep{}) scenario:
1) optimization of the effective chemical pressure in ternary hydrides \cite{DiCataldo_PRB_2021_LaBH8, DiCataldo_NPJ_2022_BaSiH, DiCataldo_PRB_2020_CaBH, Zhang_PRL_2022_LaXH, Zhang_PRL_2022_LaXH, Hilleke_arXiv_2022_chempress}, through a careful combination of guest elements in host-guest metallic superhydrides; 
and 2) doping of non-hydride covalent structures \cite{Ekimov_Nature_2004_bdopedC, Ekimov_DiamRelMat_2010_sc,Cui_JPCC_2020_XB3Si3,Flores_H2O,Pickett_PRL_2002_LiBC,Saha_PRB_2020_BC,DiCataldo_PRB_2022_XB3C3},
which exploits the large intrinsic \ep{} coupling
of covalent insulators, turned  metallic via either external or self doping, as in B-doped diamond or MgB$_2$ \cite{Kortus_PRL_2001_MgB2,Boeri_PRL_2004,Ekimov_Nature_2004_bdopedC,Strobel_APS_2022_XB3C3}. 

Both approaches present potential drawbacks: while even the most optimized ternary hydrides  seem to require synthesis pressure
of at least a few GPa \cite{DiCataldo_NPJ_2022_BaSiH}, 
the \Tc's of systems containing exclusively boron, carbon and other heavier elements are unlikely to exceed the 80 K 
threshold found in  best-case scenarios \cite{Saha_PRB_2020_BC, DiCataldo_PRB_2022_XB3C3}, 
due to the intrinsic phonon energy scales determined by the relatively large boron and carbon atomic masses.

In this paper we propose a hybrid strategy which combines the best of the two:  
doping covalent bonds (ambient pressure) in a hydrogen-rich structure
(higher \Tc). 
In particular, we show that metal borohydrides (MBH) can be turned into high-temperature conventional superconductors at 
ambient pressure, via small substitutional doping at the metal site,
which effectively transforms MBH into
highly-tunable hole-doped hydrocarbons.

MBH form a broad class of materials widely used in commercial hydrogen storage
applications, due to the high hydrogen content, and the ease of hydrogen uptake and dehydrogenation \cite{Zuttel_ChemRev_2007_MBH, Zuttel_Energies_2011_MBH}. In these compounds boron and hydrogen form quasi-molecular units arranged on open structures, 
with mono-, di- or trivalent metals ($M$) on the interstitial sites.
Our strategy to turn MBH into high-\Tc{} superconductors is quite general, and consists in replacing a small fraction of $M$ atoms with a
lower-valence atom, realising hole doping; in this work, we study the specific case of K- doping of the $\alpha$ phase of \cabh{} \cite{Chandra_ActaMat_2009_CaB2H8_phases}.

Our calculations demonstrate that substitutional K doping 
in \cabh{}
is  energetically feasible  up to at least 0.10 h$^+$/f.u.;
concentrations as low as 0.03 holes per formula unit (h$^+$/fu) are sufficient to induce superconductivity with a \Tc{} as high as 110 K. 
As MBH are commercially available materials, we expect our work to find an immediate
response from experimental researchers.

\begin{figure}[tb]
	\includegraphics[width=0.8\columnwidth]{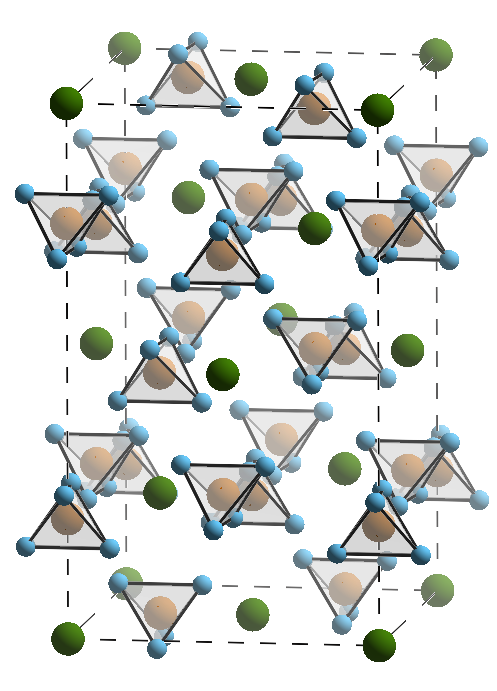}
	\caption{Crystal structure of $\alpha$-\cabh. The Ca, B, and H atoms are shown as green, orange, and blue spheres, respectively. BH$_{4}^{-}$ anions are shown as tetrahedra.}
	\label{fig:cabh_xtalstructure}
\end{figure}


$\alpha$-\cabh{} (Fig. \ref{fig:cabh_xtalstructure})
is a molecular crystal, in which boron and hydrogen
form BH$_4$ tetrahedra, and Ca occupies
interstitial sites. Ca is almost completely ionized (Ca$^{++}$), and donates charge to the BH$_{4}^{-}$ tetrahedra, which are thus not  only isostructural, but also isoelectronic to methane (C$H_4$).
The  spacing between BH$_{4}^{-}$ molecular units is quite large, about 3.5 $\AA$,
indicating extremely weak intermolecular interactions.

\begin{figure}[t]
	\includegraphics[width=1.05\columnwidth]{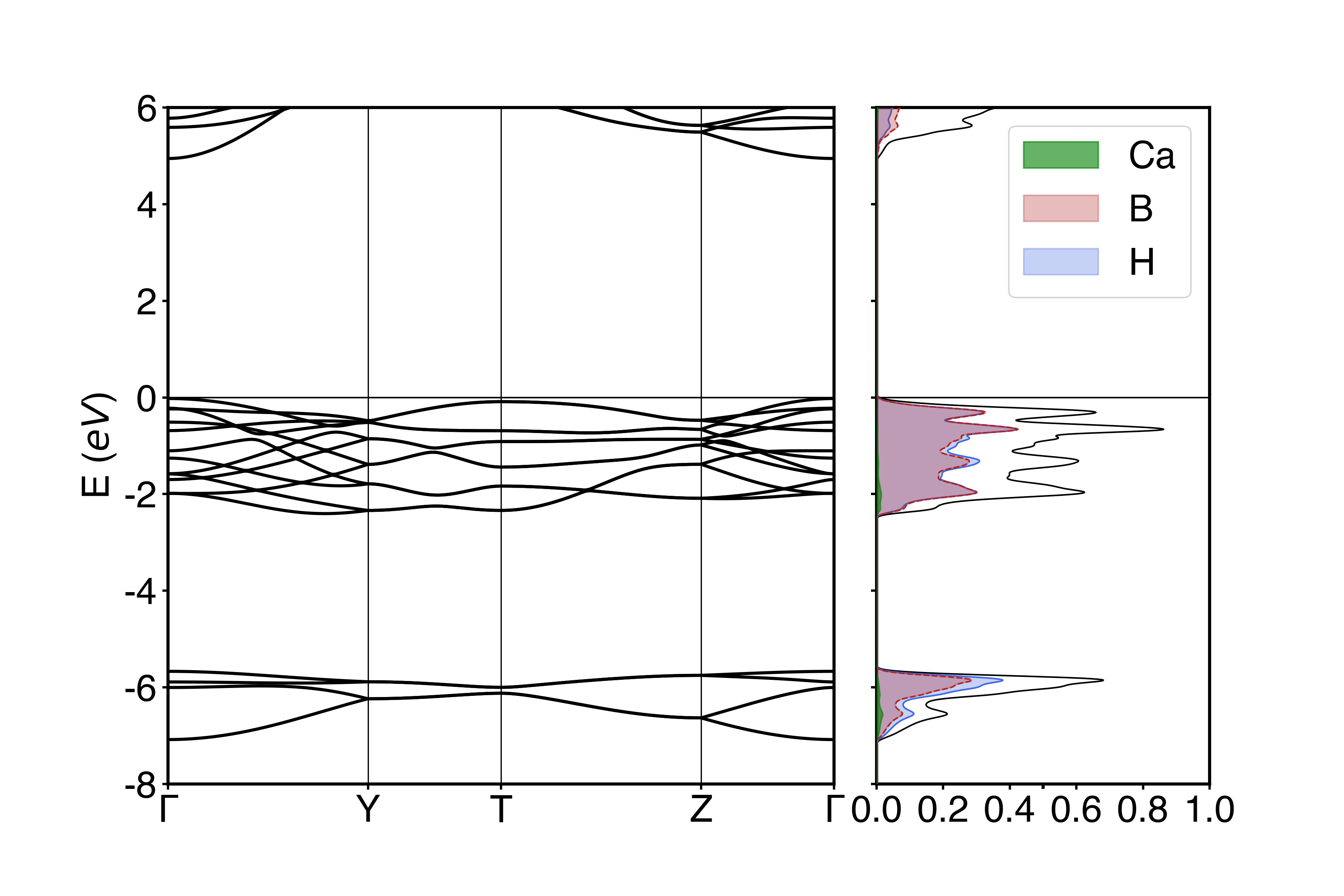}
	\caption{Electronic band structure and atom-projected DOS of \cabh{}. The energy zero is set to the valence band maximum. The DOS is in units of states $eV^{-1}atom^{-1}$.}
	\label{fig:bandstructure}
\end{figure}

Fig. \ref{fig:bandstructure} shows the electronic band structure and
the atom-projected Density of States (DOS) 
of the $\alpha$ phase of \cabh{}. 
Undoped $\alpha$-\cabh{} is an insulator, with a calculated direct band gap of 5 eV. The bands have a reduced dispersion, 
as typical of molecular crystals; the electronic DOS  exhibits  extremely sharp and narrow peaks, 
particularly near the valence band maximum (VBM). 
 Electronic states in this region have a mixed (50/50) B/H character and derive from the three-fold degenerate $1t_2$ (0 to -2 eV) and the single $2a_1$ (-6 to -8 eV) molecular $\sigma$ orbitals of BH$_4$, which are expected to couple strongly to B-H bond-stretching and bond-bending, phonons.

Due to the extremely sharp profile of the DOS, even extremely small hole dopings are sufficient to shift the Fermi energy into the large-DOS, large-\ep{} region below the VBM, which should induce high-\Tc{} conventional SC.

Hole doping in \cabh{} can be realized by substituting Ca with a monovalent atom. 
In this work, we consider K, which is the neighbour of Ca in the periodic table, and hence
has very similar size and core. Replacing a fraction $\delta$ of divalent Ca with monovalent K amounts to doping $\alpha$-\cabh{} 
with $\delta$ holes/f.u.; in an ideal rigid-band picture, these holes would form 
at the top of the valence band, turning K-doped \cabh{} into a self-doped version of methane (CH$_4$).

Due to the presence of stiff covalent bonds coupled by symmetry to
bond-stretching phonons, doped hydrocarbons have long been postulated to exhibit large \ep{} coupling; controversial
reports of high-\Tc{} superconductivity in polyacenes 
doped with alkali and alkaline earths (\textit{electron} doping)
have appeared in the early 2010's ~\cite{Devos_Lannoo_PRB1998,Kubozono_nature_2010,Subedi_PRB2011,Casula_PRB_2012}.
However, doping  hydrocarbons and related C-H systems with \textit{holes} has so far proven impossible,
as intercalation with electronegative elements (I,F, Cl) is ineffective, and doping the C-H sublattice by substitutional atoms or vacancies 
is extremely unfavorable energetically and tends to seriously disrupt the crystal and electronic structure due to the presence
of stiff covalent bonds \cite{Giustino_PRL_2010_Graphane,FloresLivas_eupj_2018_dopedpolyeth}.
 
 In K-doped \cabh{}, on the other hand, doping only involves the metal site, which is very weakly bonded
 to the rest of the structure. This should allow a convenient fine-tuning of the superconducting properties
 at a reasonable energy cost,
 without major modifications of the structure  \cite{Moussa_PRB_2008_fragments}.
 
 To substantiate our hypotheses, we computed the superconducting properties of K-doped \cabh{} for 
 various hole concentrations $\delta$, computing the isotropic \'{E}liashberg functions for various values of $\delta$ using Density Functional Perturbation Theory \cite{Baroni_RevModPhys_2001_DFPT}, and obtaining the \Tc{} by numerically solving the isotropic \'{E}liashberg equations \footnote{Calculations were performed using DFPT as implemented in Quantum Espresso. Integration of electron-phonon properties was performed on $3\times3\times3$ grid for phonons and $16\times16\times16$ grid for electrons, using a Gaussian smearing of 200 meV. Further computational details are provided in the Supplementary Material \cite{suppmat, Kresse_PRB_1996_VASP, Izumi_JAC_2008_VESTA, Baroni_RevModPhys_2001_DFPT, quantumespresso_1, quantumespresso_2, Hamann_PRB_2017_ONCV}}.
  
Doping is simulated using the virtual crystal approximation (VCA),
which amounts to replacing each Ca (pseudo)atom in the $\alpha$-\cabh{} structure with an average virtual (pseudo)atom, obtained by mixing K and
Ca in the appropriate proportions.
  
\begin{figure}[t]
	\includegraphics[width=1.0\columnwidth]{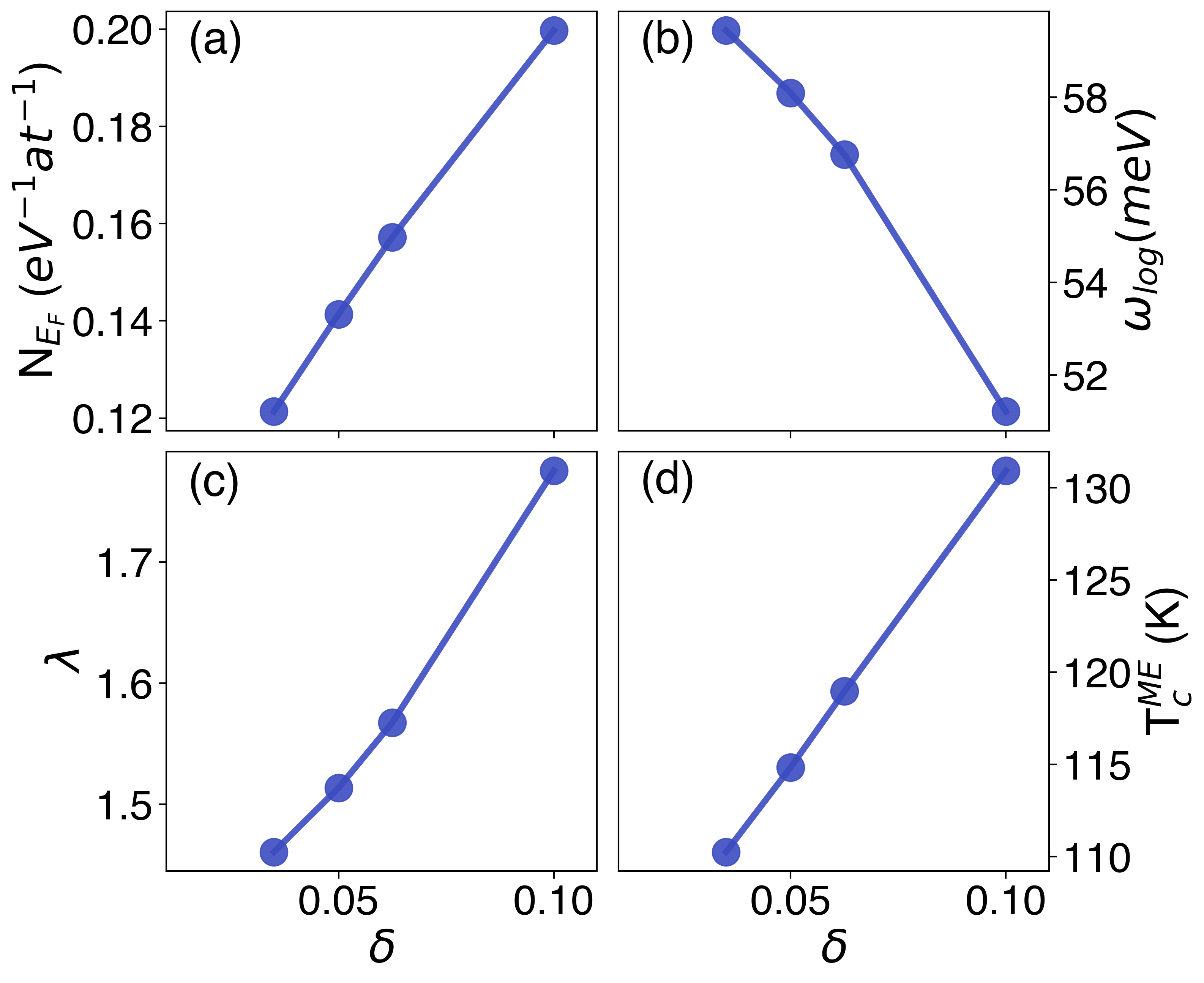}
	\caption{Electronic and superconducting properties as a function of the doping $\delta$. (a): DOS at the Fermi level, (b): logarithmic average phonon frequency $\omega_{log}$, (c): \ep{} coupling coefficient $\lambda$, (d): superconducting critical temperature \Tc{}.}
	\label{fig:Tc_doping}
\end{figure}

In Fig. \ref{fig:Tc_doping} we report a summary of the electronic and superconducting properties of K-doped \cabh{} as a function of the hole concentration $\delta$. The DOS at the Fermi level N$_{E_{F}}$ (panel (a)) rapidly increases with doping, as does the total \ep{} coupling constant $\lambda$ (panel (c)), while the average phonon frequency $\omega_{log}$ (b) slightly decreases. In the \'{E}liashberg function (shown in Fig. S2 of the Supplementary Material) almost all \ep{} coupling is concentrated in the high-energy B-H stretching and bending modes. For  $\delta$ larger than 0.10 the system develops a dynamical instability, while for values smaller than 0.03 the Fermi energy is too close to the VBM to allow a reasonable estimate of $\lambda$ and \Tc{}. 
 \Tc{} attains its maximum value of 130 K at $\delta$ = 0.10 and decreases linearly with decreasing $\delta$ down to 110 K at $\delta$=0.03;
extrapolating this trend, we can reasonably suppose that 
\Tc{}'s higher than 100 K may be achieved for dopings $\delta \gtrsim 0.01$.

The VCA has the advantage of making \Tc{} calculations
feasible even for small dopings, correctly capturing  the average effect of substitutional doping on the electronic structure, 
in particular the critical role
of electronic screening on phonon spectra and \ep{} matrix elements.~\cite{Casula_PRB_2012}
However, effects such as charge localization, deep defect levels or carrier trapping \cite{Freysoldt_RevModPhys_2014_point_defect, Flores_H2O}, which may sensibly affect the electronic structure,
require more complex approximations that can capture
depend on the local environment of the impurities.

\begin{figure}[t]
	\includegraphics[width=1.0\columnwidth]{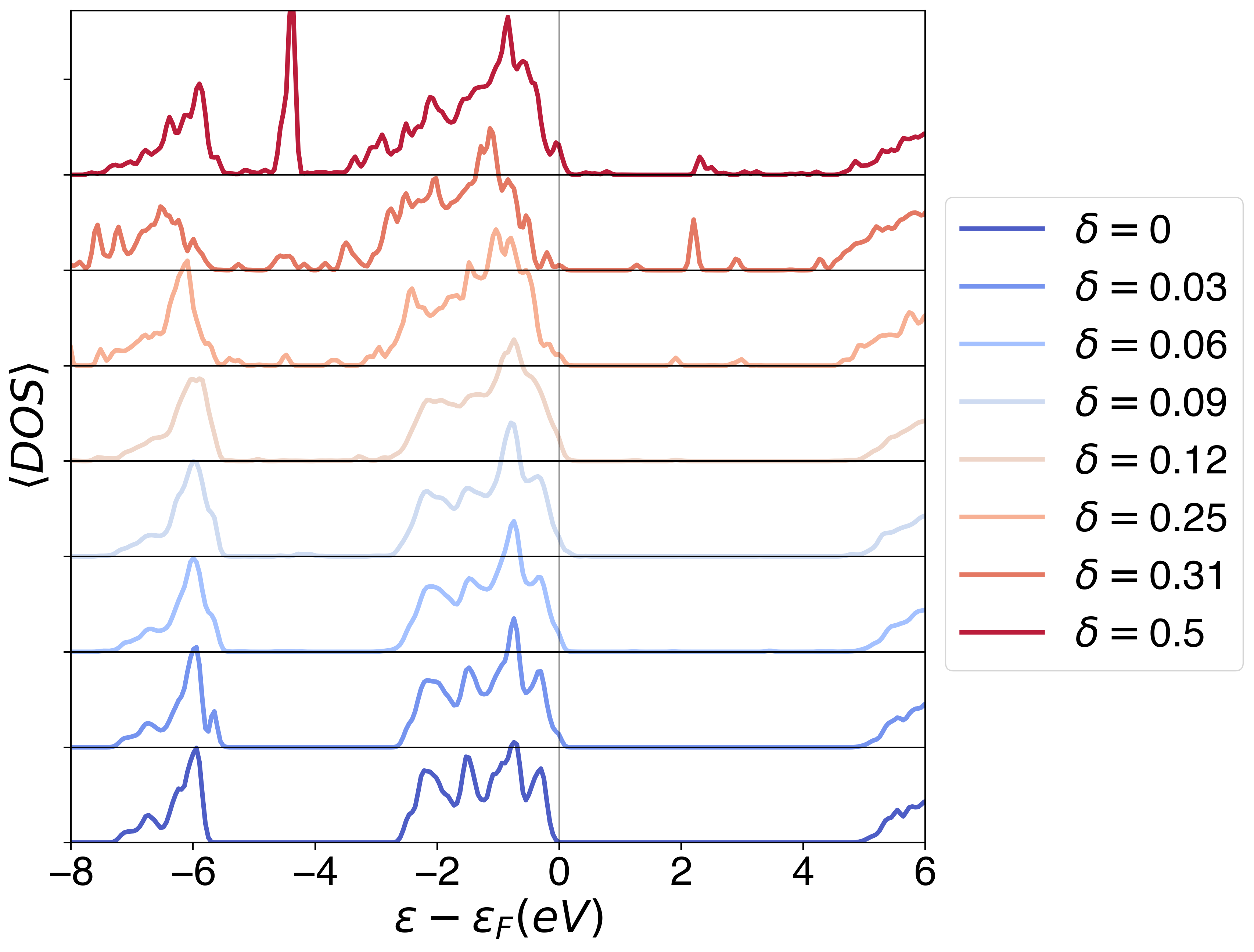}
	\caption{Density of states (DOS) of K-doped \cabh{} as a function of the hole concentration $\delta$. The Fermi energy is taken as the energy zero except for $\delta = 0$, in which the VBM is used.}
	\label{fig:DOS_doping}
\end{figure}

To simulate these effects, we developed an ad-hoc scheme,
based on averaging over random supercells.
First, we constructed a 2$\times$2$\times$2 supercell containing 32 formula units of \cabh{} (352 atoms); then, 
to simulate hole concentrations from $\delta = 0.03$ to $\delta=0.5$,
we substituted  Ca atoms with the appropriate fraction of K, 
placed at random positions; for each value of $\delta$, we generated ten supercells. These supercells were then relaxed
to minimize stress and forces, before computing the  total energies and DOS's -- Figs.~\ref{fig:DOS_doping}. Computations on the supercells were performed using the Vienna Ab-initio Simulation Package (VASP). Further details are provided in the Supplementary Materials \cite{suppmat}. 
The average DOS for each doping was then obtained by
performing a weighted average over the relative supercells,  with weights corresponding to the probability of that configuration (See SM for further details \cite{suppmat}). The average DOS's for different values of $\delta$ are shown in Fig. \ref{fig:DOS_doping}.
Although doping causes sizable modifications of the
DOS for $\delta > 0.12$, especially in the low-lying region below -6 eV, the DOS for $\delta$ up to 0.09 are essentially unchanged
compared to the undoped compound. In particular, there are 
no inter-gap states up to $\delta$ = 0.09, and the relative
weigth remains negligible up to $\delta=0.25$.

Hence, for doping of interest
the main effect of K/Ca substitution is indeed a quasi-rigid shift of the Fermi level into the valence band, well reproduced by VCA; 
in particular, the states just below the VBM, which 
participate in the SC pairing, are only weakly influenced by doping. 
This is expected, since $\sigma$ orbitals of the BH$_4^{-}$
molecular ions are only weakly affected by distortions and rearrangements of atoms in the crystal structure which do not modify the overall shape of the molecular ion itself.

\begin{figure}[t]
	\includegraphics[width=0.9\columnwidth]{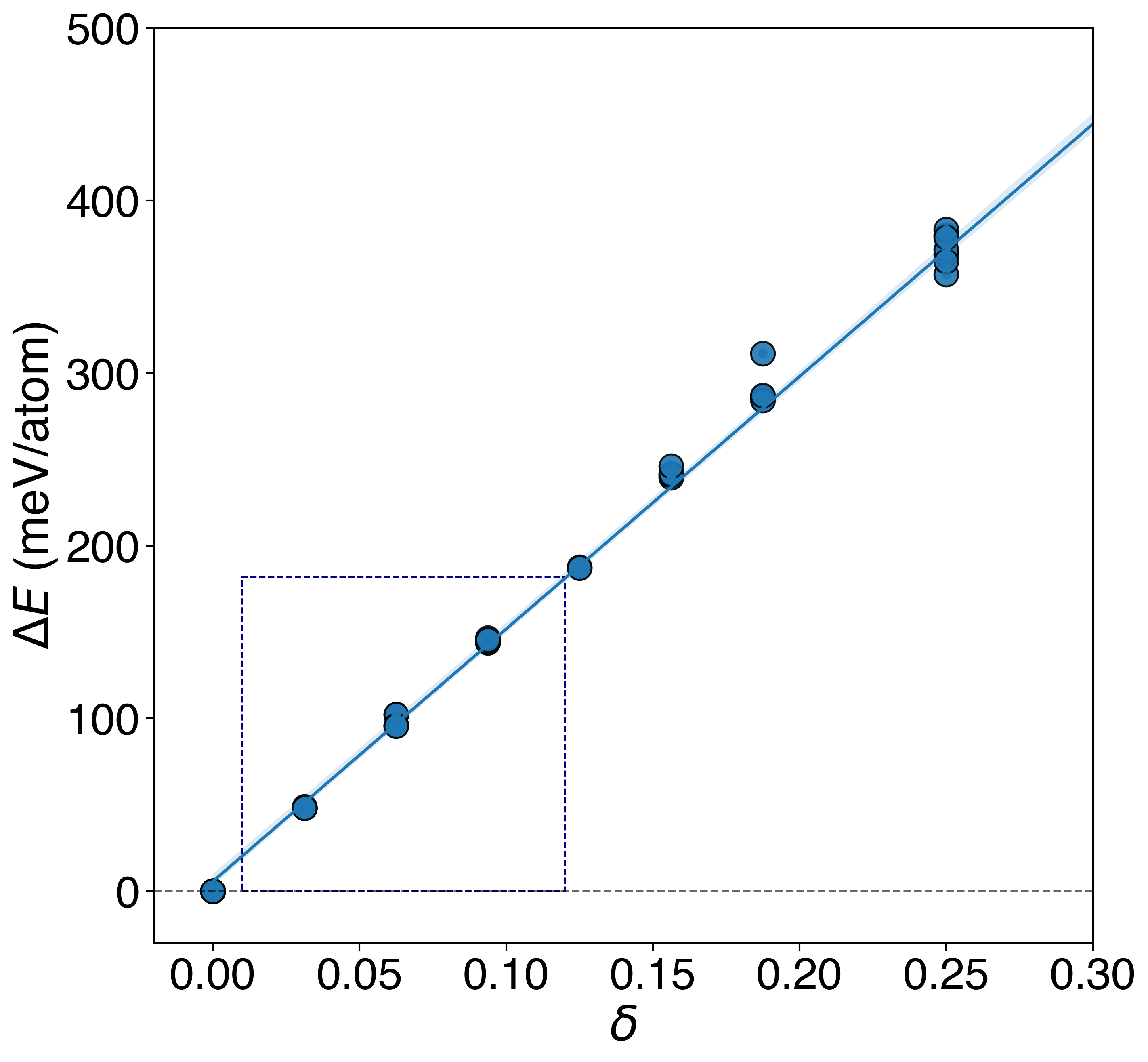}
	\caption{Formation energy $\Delta E$ as a function of hole-doping $\delta$.}
	\label{fig:dH_doping}
\end{figure}

Using supercells we can also estimate the energetic cost of K substitution into the Ca site. In Fig. \ref{fig:dH_doping} we show the formation energy of K$_{\delta}$Ca$_{1-\delta}$(BH$_4$)$_2$ with respect to decomposition into K + \cabh{} \footnote{For Ca and K we assumed a face-centered and a body-centered cubic structure, respectively}, for all configurations sampled (ten for each doping). The formation energies $\delta E$ increase linearly with doping, with little dispersion for different supercells, remaining below 200 meV/atom for $\delta$ up to 0.12. 
 
 On purely energetic grounds, these values indicate that K-doping 
 of \cabh{}
 should be experimentally feasible. However,
 experimental synthesis conditions depend on complex details of the kinetic barrier protecting the doped structure from decomposition, and on the entropy contribution to the free energy, whose evaluation
 goes well beyond the scope of this paper. 
 
Moreover, independently of their energetic cost, not all perturbations induced by doping  will have the same effect on superconductivity;
 while small distortions and rearrangements of BH$_{4}^{-}$ anions within the open $\alpha$ structure, 
 should have only minor effects on the the superconducting properties,
 dehydrogenation, which implies a weakening of the B-H bonds,
has  severe consequences, since it implies a major rearrangement of the whole electronic structure. We will therefore assume 
that the dehydrogenation energy can be used 
to estimate an effective synthesizabilty  threshold
for K-doped \cabh{}.

A hand-waving estimate  can be obtained as follows.
Assuming the measured dehydrogenation temperature
of \cabh{},  which is around 700 K \cite{Hauback_JMC_2007_CaBHx_discovery, Cho_JAC_2008_CaB2H8_dehydr, Remhof_PCCP_2016_CaB2Hx_decomposition}, to be a reasonable guess
of the kinetic barrier for dehydrogenation, doped \cabh{} should be able to withstand perturbations with a positive $\Delta E$ of the order of 60 meV/atom without decomposing. This  corresponds to a doping $\delta \sim$ of 0.06 (Fig. \ref{fig:dH_doping}), which as Fig.~\ref{fig:Tc_doping} (d) shows, largely exceeds the doping levels
required for high-\Tc{} superconductivity.
Hence, high-\Tc{} superconductivity should be observable, before
dehydrogenation sets in.

In summary, in this paper we proposed a strategy to attain high-\Tc{} conventional superconductivity in the complex bordohydride \cabh{} using substitutional doping of monovalent K on the Ca site, and substantiated it with first-principles calculations.

K-doped \cabh{} behaves essentially as hole-doped methane (CH$_4$), where the high-\Tc{} derives from a strong coupling 
between $\sigma$ bonding electronic states and bond-stretching 
phonons of the BH$_4^{-}$ molecular units. Compared to
CH$_4$, however, the big advantage of \cabh{} is that
hole doping is realized by acting on the weakly-bonded 
metal site, and not on the covalent B-H (or C-H) sublattice,
and this causes only minor disruptions of the crystal and 
electronic structure, implying an affordable energy cost.
According to our calculations, a partial replacement of 3 $\%$ Ca atoms with K atoms would have an energy cost of around 50 meV/atom, which is below the dehydrogenation threshold, and lead to an estimated \Tc{} of 110 K at ambient pressure, almost on par with the best copper-oxide superconductors. With a figure of merit $S$ between 2.8 and 3.3 \cite{Pickard_AnnRevCMP_2020_review}, doped \cabh{} is better than any other superhydride, as well as all other known ambient-pressure conventional superconductors ($S=1$ in MgB$_2$), and very close to HgBaCaCuO.

Note that the strategy proposed here is very general, and can in principle be applied to
turn  any of the many existing MBH into doped hydrocarbons, by suitable metal substitutions.
 We are strongly convinced that, if synthesized, doped metal borohydrides will represent a huge leap forward in research on high-temperature superconductors.

Given the easy commercial availability of metal borohydrides, we hope that our work will stimulate a positive response from experimentalists.

\section{Acknowledgments}
The authors warmly thank Antonio Sanna for sharing the code to solve the isotropic \'{E}liashberg equations. L.B. and S.d.C. acknowledge funding from the Austrian Science Fund (FWF) P30269-N36 and support from Fondo Ateneo-Sapienza 2017-2020. S.D.C. acknowledges computational resources from CINECA, proj. IsC90-HTS-TECH and IsC99-ACME-C, and the Vienna Scientific Cluster, proj. 71754 "TEST".

%

\end{document}